\documentclass[useAMS]{mn2e}
\usepackage{graphicx}
\usepackage{epsfig}

\def\ltsima{$\; \buildrel < \over \sim \;$}
\def\lsim{\lower.5ex\hbox{\ltsima}}
\def\gtsima{$\; \buildrel > \over \sim \;$}
\def\gsim{\lower.5ex\hbox{\gtsima}}

\newcommand{\be}{\begin{equation}}
\newcommand{\en}{\end{equation}}
\newcommand{\ergs}{\rm \ erg \; s^{-1}}

\def\msole {~M_{\odot}}

\begin{document}
\title[]{The return to quiescence of Aql X-1 following the 2010 outburst}

\author[Campana et al.]{S. Campana$^{1,}$\thanks{E-mail: sergio.campana@brera.inaf.it}, F. Brivio$^{1,2}$, N. Degenaar$^3$, 
S. Mereghetti$^4$,  R. Wijnands$^5$, \newauthor 
P. D'Avanzo$^1$, G. L. Israel$^6$, L. Stella$^6$, \\ 
$^1$ INAF - Osservatorio astronomico di Brera, Via Bianchi 46, I--23807, Merate (LC), Italy\\
$^2$ Universit\`a di Milano, Via Celoria 16, I--20133, Milano, Italy\\
$^3$ Department of Astronomy, University of Michigan, 500 Church Street, Ann Arbor, MI 48109, USA\\
$^4$ INAF - IASF Milano, Via Bassini 15, I--20133 Milano, Italy\\
$^5$ Astronomical Institute Anton Pannekoek, University of Amsterdam, Postbus 94249, 1090 GE Amsterdam, The Netherlands\\
$^6$ INAF - Osservatorio astronomico di Roma, Via Frascati 44, I--00040, Monteporzio Catone (Roma), Italy
}

\maketitle

\begin{abstract}
Aql X-1 is the most prolific low mass X--ray binary transient hosting a neutron star. In this paper we focus on the return to quiescence following the 
2010 outburst of the source. This decay was monitored thanks to 11 pointed observations taken with XMM-Newton, Chandra and Swift.
The decay from outburst to quiescence is very fast, with an exponential decay characteristic time scale of $\sim 2$ d. Once in quiescence 
the X--ray flux of Aql X-1 remained constant, with no further signs of variability or decay. The comparison with the only other well-monitored outburst from 
Aql X-1 (1997) is tail-telling. The luminosities at which the fast decay starts are fully compatible for the two outbursts, hinting at a mechanism intrinsic to the 
system and possibly related to the neutron star rotation and  magnetic field (i.e., the propeller effect). In addition, for both outbursts, the decay profiles are 
also very similar, likely resulting from the shut-off of the accretion process onto the neutron star surface. Finally, the quiescent neutron star temperatures 
at the end of the outbursts are well consistent with one another,  suggesting a hot neutron star core dominating the thermal balance. 
Small differences in the quiescent X--ray luminosity among the two outbursts can be attributed to a different level of the power law component.
\end{abstract}


\section{Introduction}

Bright low mass X--ray binary transients (LMXTs) hosting a neutron star show two main different outburst 
phenomenologies and can be broadly divided in ''quasi-persistent'' and ''classical'' transients. 
Quasi-persistent neutron star transients show prolonged active states, lasting up to several years. 
Sources like  MXB 1659--29 ($\sim 2.5$ yr activity, Wijnands et al. 2004; Cackett et al. 2008 and references therein), 
KS 1731--260 ($\sim 12.5$ yr activity, Wijnands et al. 2002; Cackett et al. 2010 and referencs therein), EXO 0748--676 
($\sim 24$ yr activity, Degenaar et al. 2009, 2011 and referencs therein), XTE J1701--462 ($\sim 1.6$ yr, Fridriksson et al. 2010) 
or HETE J1900.1--2455 (still active, Papitto et al. 2013 and reference therein) belong to this class. 
The best sampled outburst X--ray light curve pertains to XTE J1701--462. After a smooth and long outburst peak,
the X--ray luminosity decreased by a factor of $\sim 2000$ in  $\sim 13$ d, before starting a much slower and smooth decay
driven by the crust cooling of the neutron star surface (Fridriksson et al. 2010). 

The other broad class of bright transients shows outbursts with a shorter duration (weeks-months), often characterized by a  
fast rise time (a few days) and a smooth exponential decay (about one month), although many also show more complex 
outburst light curves. 
The prototype of this class is Aql X-1 (e.g. Campana, Coti Zelati \& D'Avanzo 2013).
Monitoring of its 1997 outburst with BeppoSAX  showed a steep decay when the luminosity  reached a 
level of $\sim 5\times 10^{36}$ erg s$^{-1}$ (at 4.5 kpc), which can be described by an exponential with an 
$e-$folding time of $\sim 2$ d (Campana et al. 1998a). Following this decay, Aql X-1 attained its quiescent level.
Similarly steep decays have later been observed  (even though with less details) in other sources,  such as SAX J1808.4--3658 
(Gilfanov et al. 1998; Campana, Stella \& Kennea 2008); the Rapid Burster (Masetti et al. 2000) and XTE J1709--267 
(Jonker et al. 2003; Degenaar, Wijnands \& Miller 2013a).
This behavior has been observed also in a few sources classified as faint transients\footnote{Sources in this subclass 
differ from the bright LMXTs because
they reach  fainter peak luminosities ($\sim 10^{36}$ erg s$^{-1}$)  and usually show shorter outburst durations.} 
such as IGR J00291+5934 (Jonker et al. 2005), Swift J1756.9--2508 (Krimm et al. 2007) and Swift J1749.4--2807 
(Ferrigno et al. 2011; however
one outburst of this source showed a linear decay down to quiescence, Wijnands et al. 2009; Campana 2009).  

Neutron star transients in quiescence are characterised by an X--ray spectrum made by two different components, one soft and one hard.
The soft component can be adequately fitted with a thermal model and  is usually interpreted as the result of the cooling of the neutron 
star surface, heated during repeated outbursts (deep crustal heating; Brown et al. 1998; Wijnands 2011). 
Alternatively this component could be powered by low level accretion (van Paradijs et al. 1987; Campana et al. 1998a; 
Cackett et al. 2010; Fridriksson et al. 2010).
The hard  component can be described by a power law with photon index $\Gamma\sim 1.5-2$.
Several possibilities have been suggested for its origin, including residual accretion stopped at the magnetospheric radius 
and shock from a pulsar wind  (see e.g. Campana et al. 1998b).

The cooling of the neutron star crust has been studied in detail in quasi-persistent transients, characterised by long cooling times (years),
but very little is known on the crust cooling properties of classical (shorter duration) transients. 
Recently, Degenaar et al. (2013a) monitored the return to quiescence of XTE J1709--267 with XMM-Newton observations,
finding  a very rapid  temperature decrease. 
Two other (faint) transients showed signs of cooling: IGR J17480--2446 in Terzan 5 
(Degenaar et al. 2013b) and  IGR J17494--3030 
(Armas-Padilla, Wijnands \& Degenaar 2013).

In this paper we report on the 2010 outburst decay from Aql X-1, the most prolific transient. The 2010 outburst was one of the brightest 
ever, reaching a peak luminosity of $\sim 350$ mCrab around Sep. 14, 2010. The outburst was preceded by a precursor event. It
raised to the maximum in $\sim 13$ d and then slowly decrease to quiescence in $\sim 47$ d (Campana, Coti Zelati \& D'Avanzo 2013).
The outburst decay has been followed in great detail
with 5 XMM-Newton, 4 Chandra and 2 Swift-XRT observations, monitoring the latest stages of the outburst.
In Section 2  we describe the data extraction. In Section 3 we fit the resulting spectra and compare the light curve decay with that of  the only other 
well sampled outburst from this source, which occurred in 1997. In Section 4 we discuss the results and in Section 5 we draw our conclusions.

\section{Data reduction}

\subsection{XMM-Newton}

Aql X-1 was observed five times with XMM-Newton during the decay of the 2010 outburst 
(see Table 1)\footnote{One earlier observation was carried out during the same outburst but only the RGS instruments were operated. 
Due to the limited bandpass we do not consider these data here.}.
Observations were carried out with the EPIC instrument, which consists of three CCD cameras (one pn and two MOS). 
The thin filter was used in front of all cameras for all the observations. Data were reprocessed using SAS v.11.0.
Data were filtered with the FLAG==0 option and only events with pattern $\leq4$ ($\leq12$) were retained for the pn (MOS) detector.
Data were accumulated in the 0.3--10 keV (0.5--10 keV) energy band for the pn (MOS) detector.
During all the observations the background remained low and we retained the entire observing time.
Burst mode and timing data from the pn detector and timing data from the MOS2 detector were 
extracted form a 16 pixel wide strip centred on the source (MOS1 timing data were not used because the source felt on a bad column). 
Large or full window data were extracted from a 750 (650) pixel radius region for the pn (MOS).
Count rates for all the observations were reported in Table \ref{obslog}. The background was evaluated from the same CCDin two  
regions close to Aql X-1 and free of sources and bad columns. 
The spectral channels were grouped to have at least 20 counts per  bin and {\tt rmf} and {\tt arf} files were generated using the latest calibration 
products. 

\begin{table*}
\caption{Observation log.}
\label{obslog}
\begin{center}
\begin{tabular}{ccccc}
\hline
Observation$^+$& Start date      & Duration & Obs. ID.        & Count rate\\
                               &                        & (ks)$^\#$  &                       &(c s$^{-1}$)\\
\hline
XMM\_1 (BTT)& Oct 12, 2010&  18.1       & 0085180101& $14.0\pm0.06$\\
XMM\_2 (TLT) & Oct 17, 2010&   6.1        & 0085180201&$5.48\pm0.03$\\
XMM\_3 (TLT) & Oct 25, 2010&   6.4        & 0085180301&$0.07\pm0.01$\\
XMM\_4 (FFF) & Oct 29, 2010&   7.2        & 0085180401&$0.10\pm0.01$\\
XMM\_5 (FFF) & Nov 02, 2010& 14.4       & 0085180501&$0.10\pm0.01$\\
\hline
Swift\_1          & Oct 15, 2010 & 2.6          &00030796248&$0.37\pm0.01$\\
Swift\_2          & Oct 18, 2010 & 2.8          &00030796249&$0.23\pm0.01$\\
\hline
Chandra\_1 & Oct 19, 2010 & 6.4           & 12456 &  $0.59\pm0.01$\\
Chandra\_2 & Oct 22, 2010 & 6.4           & 12457 & $0.26\pm0.01$\\
Chandra\_3 & Oct 30, 2010 & 6.4           & 12458 & $0.15\pm0.01$\\
Chandra\_4 & Nov 12, 2010& 7.0           & 12459 & $0.16\pm0.01$\\
\hline
\end{tabular}
\end{center}
{\leftline{$^+$ The observing mode of the three XMM-EPIC instruments is coded as B:burst, T: timing, L: Large window and F: full window.
Swift/}}
{\leftline{ XRT observations were obtained in Photon Counting mode. Chandra observations were obtained in Timed Exposure mode.}}
{\leftline{$^\#$ Exposure times and count rates (pile-up corrected, in the 0.5--10 keV energy range) refer to the MOS1 instrument.}}
{\leftline{$^*$ There are two additional observations with Swift that were not used in the present analysis due to the insufficient number 
of counts for}}
{\leftline{a spectrum.}}
\end{table*}

\subsection{Swift}

Swift observed Aql X-1 several times during the 2010 outburst, however only two observations provided a sufficiently high number of 
counts to perform an accurate spectral analysis (see Table \ref{obslog}). Both observations were carried out in photon counting (PC) mode with a count
rate sufficiently low that no pile-up problems were present. The data were reprocessed using {\tt xrtpipeline} v.0.12.6 and 
filtered with the standard grade scheme, accepting photons with grade in the 0--12 interval. 
The source counts were extracted from a circular region with radius of 30 and 20 pixels for the two observations, respectively.
The background data were extracted from an annulus centred on source with inner and outer radii of 80 and 110 pixels and 
70 and 90 pixels, respectively. The {\tt arf} files were generated using the task {\tt xrtmkarf} and the appropriate exposure maps. 
The spectra were binned  with at least 20 photons per  energy bin in the 0.3--10 keV energy band.

\subsection{Chandra}

Chandra observed Aql X-1 four times during its return to quiescence with the ACIS-S detector in time exposed mode. 
The exposure time was set for all the observations to 0.44 s per frame (1/8 subarray), in order to mitigate possible pile-up 
problems (below $10\%$ level). In the first observation we find signs of a mild pile-up.
For this observation in the spectral fits we included a pile-up model.
Chandra spectra were extracted using the {\tt ciao} (4.5) task {\tt specextract}. A source region of 8 pixels radius 
was used and 4 background regions (radius 15 pixels) surrounding Aql X-1 were selected, avoiding contaminating sources and on the same CCD.
Spectra were binned to have at least 20 counts per  bin and the analysis was carried out in the 0.5--10 keV energy range.
The appropriate response files were generated using the latest calibration release. 

\begin{figure}
\begin{center}
\includegraphics[width=6.2cm,angle=-90]{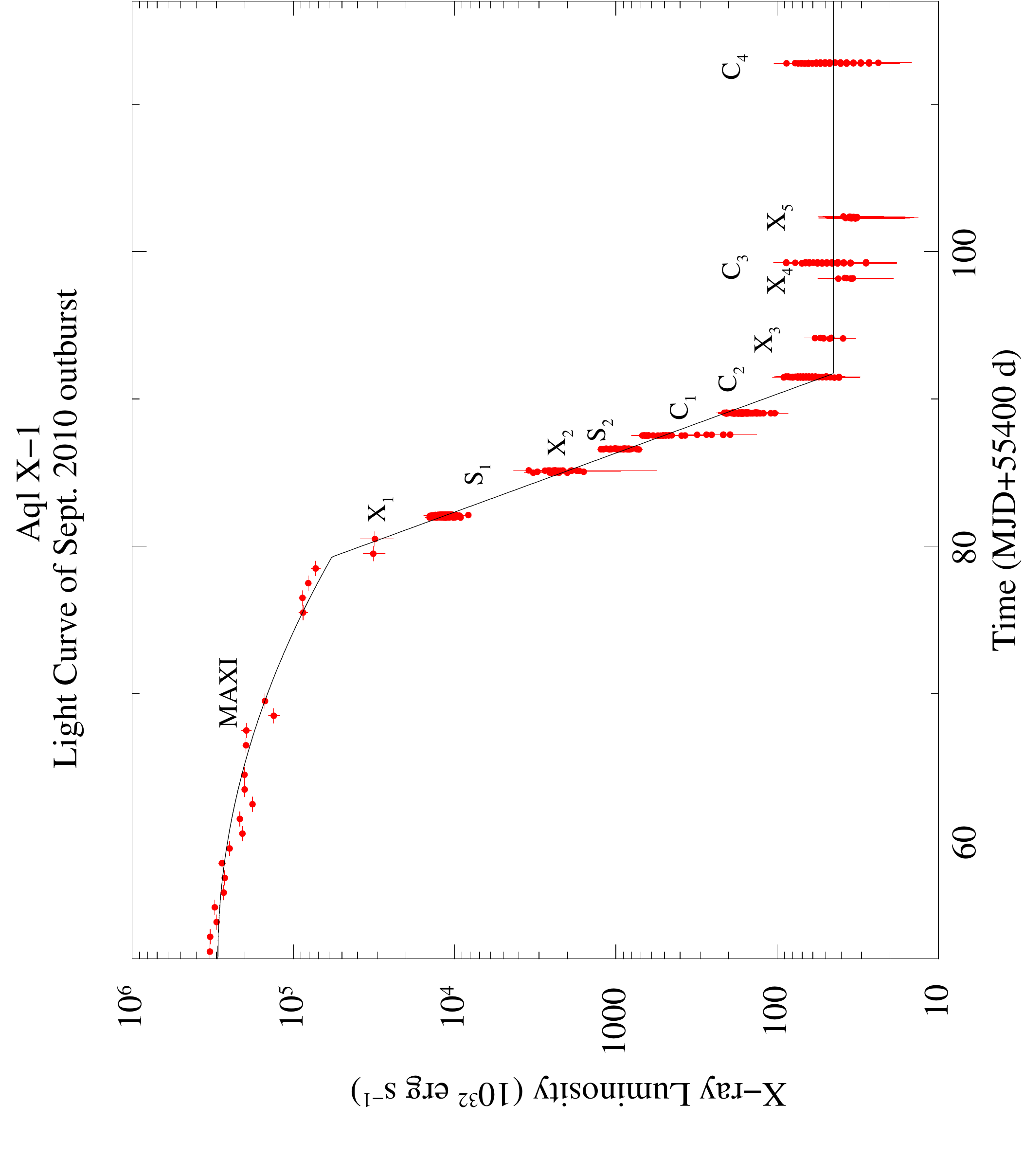}
\end{center}
\caption{X--ray luminosity light curve of the 2010 outburst from Aql X-1 as observed by MAXI, XMM-Newton, Swift and Chandra.
Count rates were converted into luminosities based on the results of the spectral fits and a distance of 4.5 kpc. 
The MAXI light curve is binned to 1 d, the focussing instruments light curves are binned to 50 s.}
\label{2010curve}
\vskip -0.1truecm
\end{figure}

\begin{figure}
\begin{center}
\includegraphics[width=6.2cm,angle=-90]{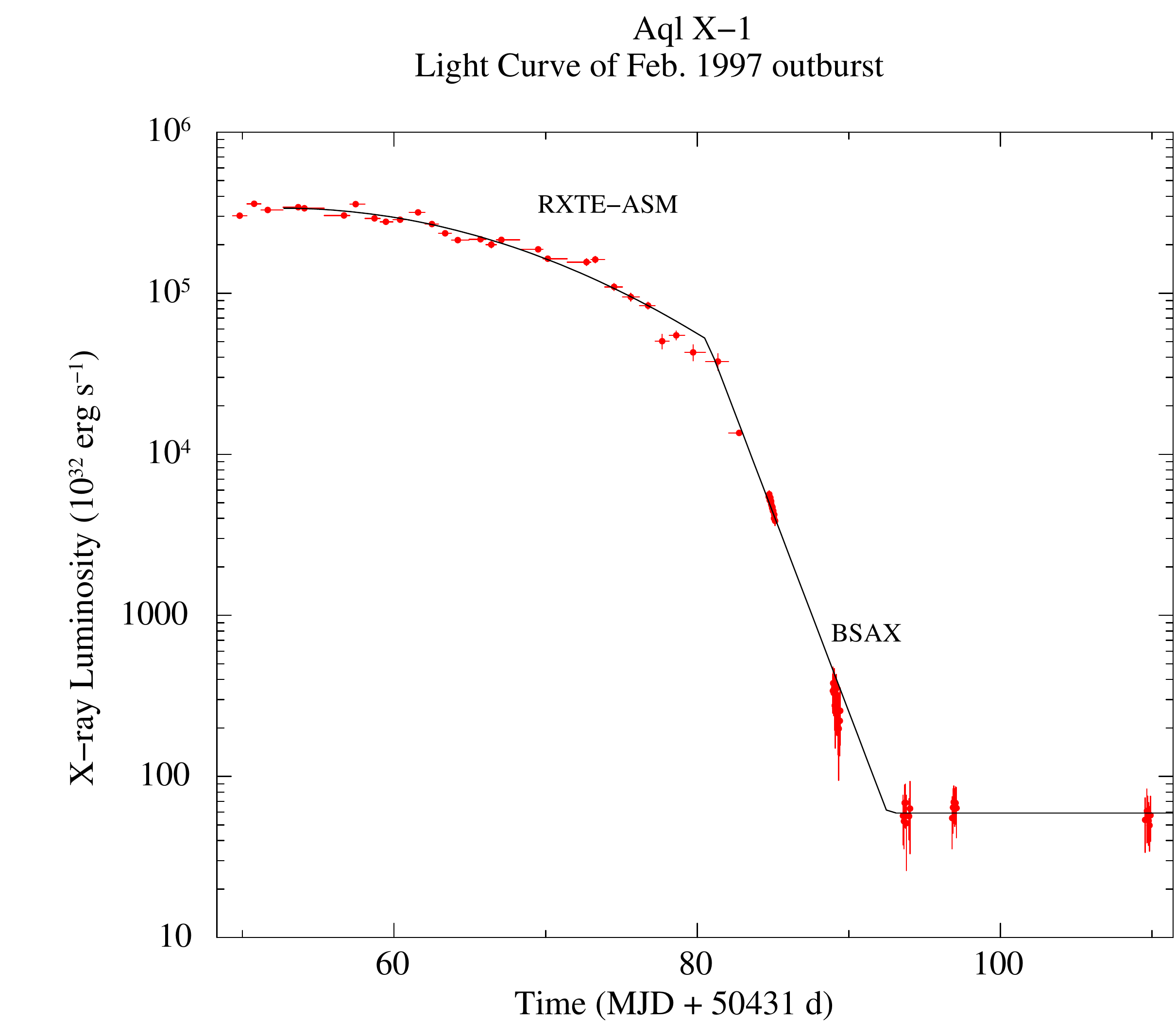}
\end{center}
\caption{X--ray luminosity light curve evolution of the 1997 outburst from Aql X-1 as observed by RossiXTE and BeppoSAX.  
Count rates were converted into luminosities based on the results of the spectral fits and a distance of 4.5 kpc.}
\label{1997curve}
\vskip -0.1truecm
\end{figure}

\section{Data analysis}

Since the XMM-Newton data are those with the highest signal to noise ratio, we first used them  for selecting the best spectral model.
We modeled the Galactic absorption with the {\tt TBABS} model (Wilms, Allen \&McCray  2000; using {\tt vern} cross-sections and {\tt wilm} 
solar abundances). We tried several single component models
never finding an acceptable description of the spectra.
A composite model made of a power law plus a softer component modeled as either a black-body, a disc model or a 
neutron star emission model can satisfactorily describe the spectra. However, fits with the disc emission model require
an absorption larger than that inferred from the  optical reddening  (Chevalier et al. 1999).
Using a black body for the soft component results in an emission radius of only a few km, much smaller than the star radius. This is 
usually interpreted as the need for a neutron star atmosphere model which provides more consistent radii (e.g. Zavlin et al. 1996).
For these reasons we used in the following the neutron star atmosphere model {\tt NSATMOS} 
(Heinke et al. 2006) for the soft component. 

\begin{table*}
\caption{Spectral fits of all the observations.}
\label{fit}
\begin{center}
\begin{tabular}{ccccccc}
\hline
Observation& Column density           &Power law                       & NS Temperature  &NS norm.$^\#$& PL fraction  & $\chi^2_{\rm red}$\\
                      &($10^{21}$ cm$^{-2}$)&photon index                  &  (eV)                       &                          &(0.5-10 keV)& (dof)\\
\hline
XMM\_1       &$5.4\pm0.1$                 &$2.06^{+0.02}_{-0.01}$&$<154$                  &$\leq1.0$         & $100\%$  &0.93 (1086)\\
Swift\_1        &$7.1\pm1.3$                &$2.38^{+0.21}_{-0.19}$&$<270$                    &$\leq1.0$          & $87\%$      &0.89 (39)\\
XMM\_2       &$3.4\pm0.1$                 &$1.16^{+0.16}_{-0.10}$&$268^{+2}_{-4}$  &$1.0_{-0.1}$& $48\%$       &1.04 (845)\\
Swift\_2        &$4.9\pm1.0$                &$0.96^{+1.26}_{-0.99}$&$249^{+24}_{-32}$&$1.0_{-0.3}$& $41\%$      &0.91 (25)\\
Chandra\_1 &$4.7\pm0.4$                &$1.26^{+0.59}_{-0.63}$&$212^{+20}_{-10}$&$1.0_{-0.4}$& $48\%$       &0.95 (120)\\
Chandra\_2 &$4.8\pm0.3$                &$0.70^{+0.74}_{-1.03}$&$180^{+17}_{-4}$  &$1.0_{-0.3}$& $36\%$       &0.81 (55)\\
XMM\_3       &$3.5\pm0.5$                 &$1.79^{+0.75}_{-0.66}$&$151^{+6}_{-12}$&$1.0_{-0.6}$& $41\%$       & 0.94 (136)\\
XMM\_4       &$4.5\pm0.3$                 &$0.93^{+0.64}_{-0.63}$&$158^{+2}_{-3}$  &$1.0_{-0.3}$& $15\%$       &1.05 (120)\\
Chandra\_3 &$4.3\pm0.9$                &$0.49^{+1.61}_{-2.14}$&$153^{+35}_{-10}$&$1.0_{-0.7}$& $31\%$       &1.29 (33)\\
XMM\_5       &$4.3\pm0.2$                 &$1.29^{+0.35}_{-0.35}$&$153^{+2}_{-2}$  &$1.0_{-0.2}$& $21\%$       &0.99 (212)\\
Chandra\_4 &$4.5\pm0.6$                &$0.97^{+1.45}_{-1.61}$&$157^{+23}_{-23}$&$1.0_{-0.5}$& $30\%$       &1.39 (35)\\
\hline
\end{tabular}
\end{center}
{\leftline{Errors were computed using $\Delta \chi^2=2.71$. We adopted the {\tt tbabs} absorption model and 
the {\tt nsatmos} neutron star atmosphere model, while}}
{\leftline{keeping fixed the neutron star mass at $1.4\msole$ and radius at 
10 km. We adopted a source distance of 4.5 kpc.}}
{\leftline{Chandra\_1 data were affected by pile-up. We fitted the data by including a pile-up model within XSPEC. The $\alpha$ parameter 
was $0.4\pm0.3$.}}
{\leftline{$^\#$ Initially the neutron star atmosphere model normalisation has been fixed to one. In a second step we computed the 
error on it, leaving }}
{\leftline{the parameter free to vary up to 1.}}
\end{table*}

The absorbed  power law plus NSATMOS model provides a good description of the spectra obtained with the different instruments during the Aql X-1  
outburst decay (see Table \ref{fit}). However, the first two observations (XMM\_1 and Swift\_1) yield a higher column density than the others (and larger
than the usual value found in quiescence) and the photon index was close to 2, at variance with that found in all other 
observations. In addition, in the first XMM-Newton observation 
the temperature of the neutron star atmosphere was very low (actually the thermal component is not required by the data). 
This might indicate that a spectral model more typical of the outburst phase is necessary for these two observations.
Therefore, we fit these spectra with an absorbed cut-off power law plus  atmosphere model.
In this way we obtain a harder photon index and  more reasonable values for the column density and for the temperature
(see the best fit parameters  in Table \ref{fitcut}).

\begin{table*}
\caption{Spectral fits for the first XMM-Newton and Swift observations.}
\label{fitcut}
\begin{center}
\begin{tabular}{ccccccc}
\hline
Observation& Column density           &Power law                       & Cut-off energy & NS Temperature  & NS norm.     & $\chi^2_{\rm red}$\\
                      &($10^{21}$ cm$^{-2}$)&photon index                  &  (keV)                &   (eV)                       &                      & (dof)\\
\hline
XMM\_1       &$4.8\pm0.2$                &$1.75^{+0.10}_{-0.07}$&$15^{+7}_{-3}$   &$>259$                  &$1.0_{-0.60}$& 0.91 (1084)\\
Swift\_1        &$6.3\pm2.4$                &$2.04^{+0.53}_{-1.58}$&$>2$                     &$<270$                 &$\leq1.0$      & 0.91 (38)\\
\hline
\end{tabular}
\end{center}
{\leftline{Errors were computed using $\Delta \chi^2=2.71$. We adopted the {\tt tbabs} absorption model and 
the {\tt nsatmos} neutron star atmosphere model,}}
{\leftline{keeping fixed the neutron star mass at $1.4\msole$ and the radius at 10 km. We adopted a source distance of 4.5 kpc.}}
\end{table*}

\begin{figure}
\begin{center}
\includegraphics[width=5.6cm,angle=-90]{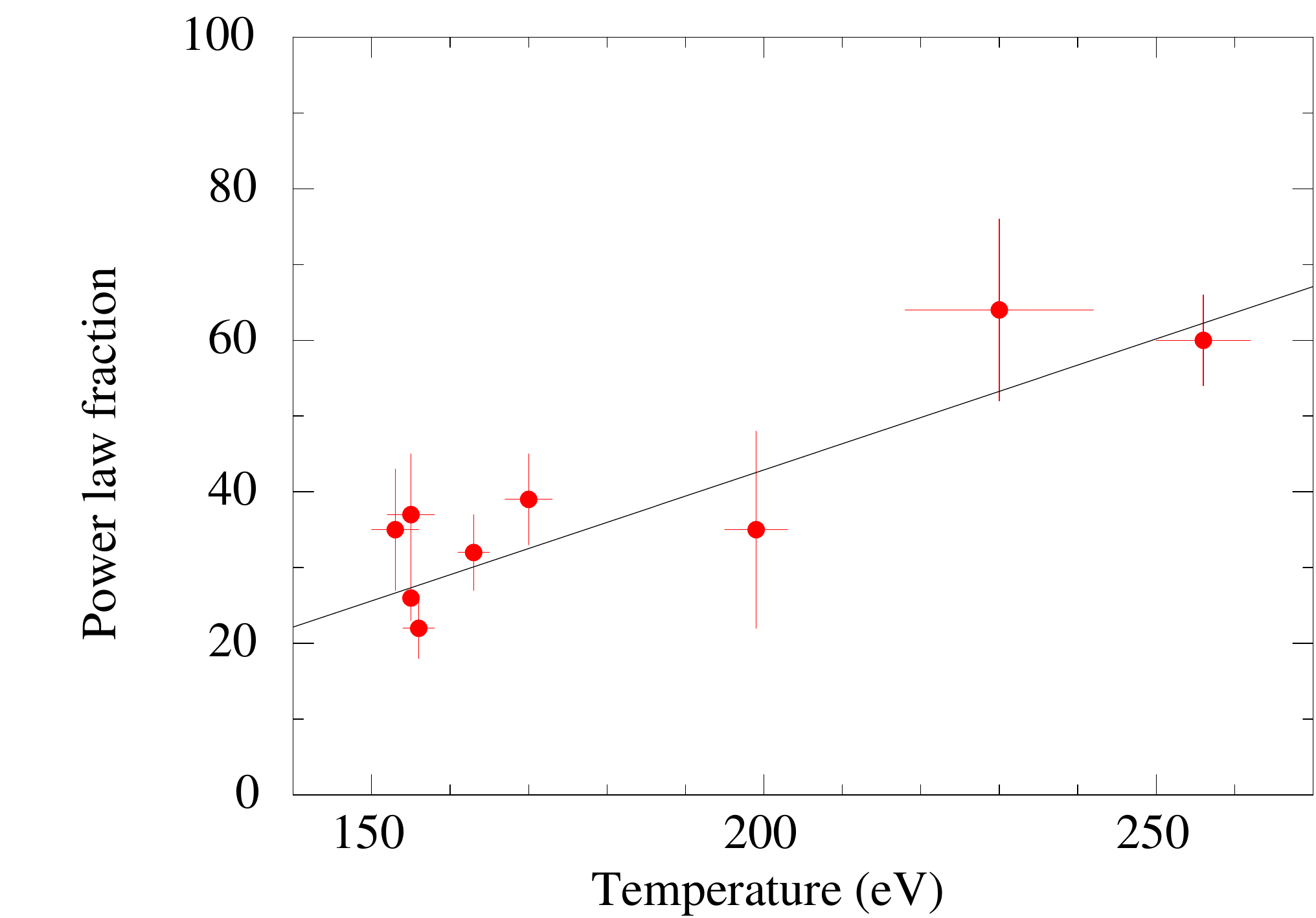}
\end{center}
\caption{Correlation between the power law fraction and the neutron star atmosphere temperature during the 2010 outburst based 
on spectral fits, taken from Table \ref{fitall2}.}
\label{tpowfrac}
\vskip -0.1truecm
\end{figure}

To characterise in greater depth the variability,
we first fit together a set of spectra to see if they can be described by the same column density
(in doing this we excluded XMM\_1, Swift\_1 spectra because they are characterised by a different spectral model, and 
Chandra\_1 spectrum, whose fit includes a pile-up correction model too).
The overall fit was good with with a reduced $\chi^2_{\rm red}=1.06$ (1524 degrees of freedom, dof) and 
null hypothesis probability (nhp) of 0.04. Results of the fit are reported in Table 
\ref{fitall}. In Table 3 we also include the fits for XMM\_1, Swift\_1 spectra, obtained by fixing the column density to 
the best common fit found before.
We then tried to group another spectral parameter. If we fit all the spectra in Table \ref{fitall} by keeping the same neutron star temperature,
we end up with a $\chi^2_{\rm red}=1.29$ (1531 dof), with a  nhp of $\sim 10^{-13}$. 
If, instead, we keep the same power law photon index we obtain a good fit with $\chi^2_{\rm red}=1.08$ (1531 dof, nhp=0.02).
As a further investigation, we computed the power law fraction to the total flux and found that it is positively correlated 
to the neutron star temperature, at least during the decay phase (see Fig. \ref{tpowfrac}).

\subsection{Light curve evolution}

Based on the spectral fits reported in Section 3 we derived the evolution of the 
outburst decay  shown in  Fig. \ref{2010curve}, where we plot the 
0.5--10 keV (unabsorbed) luminosity   (we adopt a  distance of 4.5 kpc, Galloway et al. 2008). 
To illustrate the initial part of the outburst we used data from the MAXI satellite\footnote{The light curve has been obtained from 
http://maxi.riken.jp} (Matsuoka et al. 2009),
converted into luminosity assuming a Crab-like spectrum and a 
column density of $4\times 10^{21}$ cm$^{-2}$. The overall light curve can be divided into three parts: a shallow decay followed by a 
steep decrease spanning three orders of magnitude in $\sim 10$ d and finally a  steady level. 
We modeled these three time intervals  with a Gaussian, an exponential and a constant, respectively starting from the outburst peak.
The Gaussian width ($\sigma$) was $15\pm1$ d and the steep decay started before the first XMM-Newton observation (around MJD 55479) and
had  an $e-$folding time of $1.7\pm0.1$ d. The luminosity at the knee was $5.9^{+0.2}_{-0.8}\times 10^{36}$ 
erg s$^{-1}$ (the error has been computed by propagating the error on the knee time into the luminosity curve). 
Quiescence started at MJD 55491, reaching a level of  $4.5\times 10^{33}$ erg s$^{-1}$. This was computed by 
assuming the same spectrum for the last four observations (2 XMM-Newton and 2 Chandra). The fit was fairly good with 
$\chi^2_{\rm red}=1.1$ (418 dof). The best fit values were $N_H=(4.4\pm0.1)\times 10^{21}$ cm$^{-2}$, $T=154\pm2$ eV and
$\Gamma=1.3\pm0.3$. The power law comprised $\sim 20\%$ of the total quiescent flux.

To compare the outburst light curve  with that observed in  1997 with  BeppoSAX and RossiXTE (Campana et al. 1998b),
we reanalysed the data from these satellites using  the same spectral models and distance adopted for the 2010 observations.
The 1997 outburst is the only one with an adequately sampled decay light curve.
As shown in   Fig. \ref{1997curve}, the temporal evolution of the  1997 outburst was very  similar to that of the 2010 outburst.
It can be described by  a Gaussian with $\sigma$ = $15\pm1$ d followed by an exponential decay with $e-$folding time
of $1.8\pm0.1$ d (starting at MJD 50513). The luminosity at the knee was $5.2^{+0.2}_{-0.9}\times 10^{36}$ 
erg s$^{-1}$. 
Considering  the uncertainties involved in the count rate to flux conversion 
and on the light curve modeling, this value is consistent  with that observed in 2010. 

The quiescence after the 1997 outburst started at MJD 50521, at a level of  $5.9\times 10^{33}$ erg s$^{-1}$ \footnote{This luminosity is slightly different
from that reported in  Campana et al. (1998b) because we  adopted a neutron star atmosphere model instead of a black-body,  a column
density fixed to $4.4\times 10^{21}$ cm$^{-2}$, and a distance of 4.5 kpc.},
and the spectral analysis gave  a neutron star temperature of $159^{+10}_{-6}$ eV,  similar to that observed in 2010. 
The  higher quiescent luminosity (a factor 1.3 compared to that of 2010), can be accounted by the power law  
component ($\Gamma=1.4\pm0.6$), which comprised $\sim 40\%$ of the total quiescent flux,
being the temperatures of the two quiescent states comparable.
Campana et al. (2013) estimated the energetics of the  two outbursts by integrating the overall light curves (as observed by 
by RXTE/ASM in  1997) and  MAXI (for the 2010 outburst), obtaining  $\sim 2.2\times 10^{43}$ erg and $\sim 2.8\times 10^{43}$ erg, respectively.
%
%
%
%
%
The ratio of the outbursts' fluences (2010/1997) is a factor $\sim 1.3$.  
The main parameters derived for the two outbursts are summarised in Table 6.

\begin{table*}
\caption{Spectral fits of all the observations with the same column density for all the observations.}
\label{fitall}
\begin{center}
\begin{tabular}{cccccc}
\hline
Observation& Column density           &Power law                       & NS Temperature  &Unabs. Luminosity & $\chi^2_{\rm red}$\\
                      &($10^{21}$ cm$^{-2}$)&photon index                  &  (eV)                       & (0.5--10 keV cgs)   &   (dof)\\
\hline
XMM\_1       &fixed                               &$1.46^{+0.03}_{-0.03}$&$>270$                  &$1.0\times10^{36}$& 0.91 (1085) \\
Swift\_1        &fixed                              &$1.19^{+0.34}_{-0.66}$&$8-270$                   &$1.9\times10^{35}$& 1.05 (38) \\
XMM\_2       &$4.0\pm0.1$                 &$1.83^{+0.07}_{-0.09}$&$258^{+5}_{-5}$  &$7.4\times10^{34}$& 1.06 (1524)\\
Swift\_2        &$-$                                 &$0.99^{+0.73}_{-0.45}$&$253^{+11}_{-22}$&$4.2\times10^{34}$&\\
Chandra\_1 &fixed                              &$1.05^{+0.33}_{-0.42}$&$208^{+3}_{-4}$    &$3.8\times10^{33}$& 1.16 (122)\\
Chandra\_2 &$-$                                 &$1.02^{+0.57}_{-0.54}$&$177^{+3}_{-5}$    &$7.1\times10^{33}$&\\
XMM\_3       &$-$                                  &$2.64^{+0.15}_{-0.29}$&$149^{+8}_{-7}$  &$5.3\times10^{33}$&  \\
XMM\_4       &$-$                                  &$0.51^{+0.80}_{-0.72}$&$160^{+1}_{-2}$  &$3.9\times10^{33}$&\\
Chandra\_3 &$-$                                 &$0.36^{+1.39}_{-1.08}$&$160^{+3}_{-6}$    &$4.7\times10^{33}$&\\
XMM\_5       &$-$                                  &$1.18^{+0.38}_{-0.33}$&$159^{+1}_{-2}$  &$3.8\times10^{33}$&\\
Chandra\_4 &$-$                                 &$1.14^{+1.22}_{-1.31}$&$160^{+4}_{-14}$ &$4.6\times10^{33}$&\\
\hline
\end{tabular}
\leftline{XMM\_1 was fitted with a cut-off power law with $E_c=7.5^{+0.7}_{-0.6}$ keV.}
\end{center}
\end{table*}

\begin{table*}
\caption{Spectral fits of all the observations with the same column density and power law index for all the observations.}
\label{fitall2}
\begin{center}
\begin{tabular}{cccccc}
\hline
Observation& Column density           &Power law                       & NS Temperature  &Power law  & $\chi^2_{\rm red}$\\
                      &($10^{21}$ cm$^{-2}$)&photon index                  &  (eV)                       & fraction      &   (dof)\\
\hline
XMM\_2       &$3.8\pm0.1$                 &$1.70^{+0.07}_{-0.13}$&$265^{+6}_{-5}$  &$60\pm6\%$        & 1.08 (1531)\\
Swift\_2        &$-$                                 &$-$                                    &$230^{+12}_{-12}$&$62\pm12\%$       & \\
Chandra\_1 &fixed                              &fixed                                  &$199^{+3}_{-4}$  &$35\pm13\%$     & 1.46 (123)\\
Chandra\_2 &$-$                                 &$-$                                    &$170^{+3}_{-3}$  &$39\pm6\%$        &\\
XMM\_3       &$-$                                  &$-$                                    &$163^{+2}_{-2}$  &$32\pm5\%$        & \\
XMM\_4       &$-$                                  &$-$                                    &$156^{+2}_{-2}$  &$22\pm4\%$        &\\
Chandra\_3 &$-$                                 &$-$                                    &$153^{+3}_{-3}$  &$35\pm8\%$        &\\
XMM\_5       &$-$                                  &$-$                                    &$155^{+1}_{-1}$  &$26\pm3\%$        &\\
Chandra\_4 &$-$                                 &$-$                                    &$155^{+3}_{-3}$  &$37\pm8\%$        &\\
\hline
\end{tabular}
\leftline{XMM\_1 and Swift\_1 were fitted with a cut-off power law.}
\end{center}
\end{table*}

\subsection{Temperature evolution}

\begin{figure}
\begin{center}
\includegraphics[width=7cm,angle=-90]{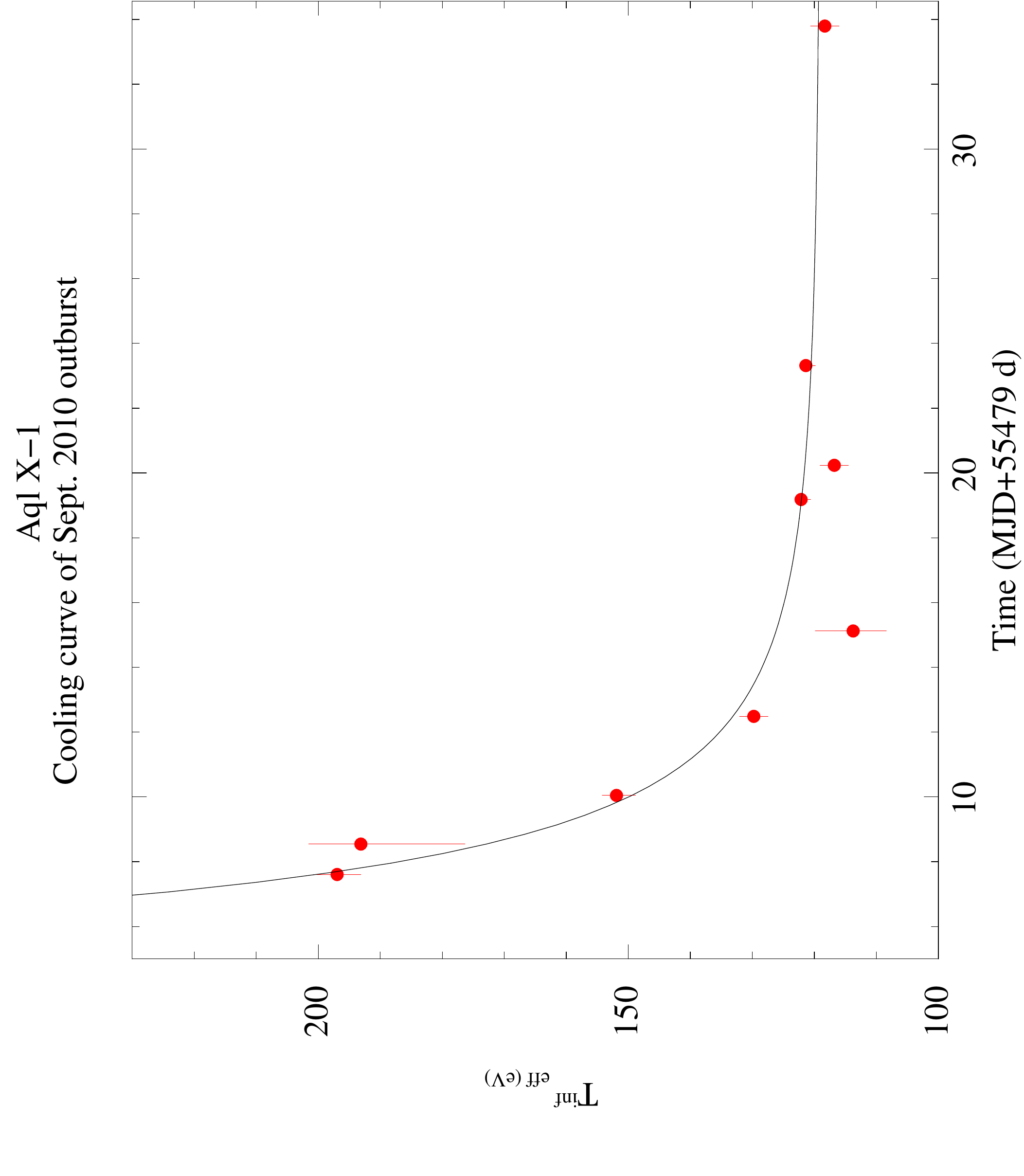}
\end{center}
\caption{Temperature evolution of the neutron star component during the 2010 outburst, based on spectral modeling with column density and 
power law spectral slope tied across all the observations. Temperature values are taken from Table \ref{fitall2} and corrected for the gravitational 
redshift.}
\label{temp}
\vskip -0.1truecm
\end{figure}

The temperatures derived with the spectral analysis described above allow us to derive a thermal evolution curve for the outburst decay.
The derived temperatures were first converted into temperatures observed by a distant observer as 
$k\,T^\infty=k\,T/(1+z)$, where $1+z=(1-R_s/R)^{-1/2}=1.31$ is the gravitational redshift factor for $M=1.4\msole$ and $R=10$ km
(with $R_s=2\,G\,M/c^2$ being the Schwarzschild radius, $G$ the gravitational constant and $c$ the speed of light).
Cooling curves are usually measured from the end of the outburst, which in our case cannot be assessed easily. 
We decided to start from the knee marking  the onset of the steep decay (MJD 55479).
The resulting temporal evolution of the neutron star temperature is shown in Fig. \ref{temp}.  
We tried to fit it with either a power law ($a\ t^{-\alpha}+b$) or
an exponential decay ($a \ \exp{(-t/\tau)}+b$). In both cases we included baseline constant level ($b$) in the fit.
The power law fit gives a reduced $\chi^2_{\rm red}=3.3$ with 6 
dof  with $\alpha=3.5\pm0.4$ and $b=119\pm2$ eV.
A better description of the data is obtained with the exponential decay, which gives 
$\tau=2.5\pm0.3$ d and $b=121\pm1$ eV ($\chi^2_{\rm red}=2.6$ for 6 dof).
These values are markedly different from what is usually found for quasi-persistent LMXBs,
where the fit with an exponential decay returns an $e-$folding time scale in the range 200--800 d (Degenaar et al. 2013b and
references therein). 
On an observational basis, it is apparent that the decay of Aql X-1 is much faster/steeper.

\begin{table}
\caption{Comparison among the 2010 and 1997 Aql X-1 outbursts}
\label{compa}
\begin{center}
\begin{tabular}{lcc}
\hline
                                                       & 2010                                                   & 1997 \\
\hline
Total energetics (erg)                & $\sim 2.8\times 10^{43}$                & $\sim 2.2\times 10^{43}$ \\
Duration  (d)                                & $\sim 60$                                           & $\sim 90$\\
Knee Lum. (erg s$^{-1}$)          &$5.9^{+0.2}_{-0.8}\times 10^{36}$& $5.2^{+0.2}_{-0.9}\times 10^{36}$\\
Decay $e-$fold time (d)             & $1.7\pm0.1$                                     & $1.8\pm0.1$ \\
Quiescent Lum. ($10^{33}$ erg s$^{-1}$) & $(4.5\pm0.2)$              & $(5.9\pm0.9)$ \\
Quiescent temperature (eV)     &  $158^{+1}_{-1}$                             & $159^{+10}_{-6}$ \\
\hline
\end{tabular}
\end{center}
\end{table}

\section{Discussion}

\subsection{Luminosity drop off}

The steep flux decay observed in the outburst of  X--ray transients is still of unknown origin.
It is important to underline that such a steep decay occurs in a number of different systems.
It has been observed in neutron star transients but also in X--ray binaries hosting a black hole (e.g. Homan et al. 2013, see also 
Shahbaz et al. 1998), as well 
as in cataclysmic variables (WZ Sge type; Kato et al. 2004). Given its ubiquity, it could be related to the  physics of the accretion disc 
rather than to the interaction of the magnetic field with the accreting matter  
(e.g. centrifugal inhibition of accretion, a.k.a. propeller effect). The disc instability model (DIM, Lasota 2001) 
is the leading model to explain LMXT (as well as dwarf novae) outbursts: an accretion disc is stable if hydrogen is everywhere 
ionised. If instead its temperature is low enough for hydrogen to recombine, the disc becomes thermally and viscously unstable. 
In this regime the disc oscillates between a hot, ionised state (outburst) and a cold, neutral state (quiescence), due to the strong 
dependence of the opacity on temperature when hydrogen is partially ionised (Lasota 2001).

DIM predicts  exponential-like decays as long as the outer disc temperature is higher than 
a critical temperature; it is expected to revert to a linear decay thereafter. This transition in the X--ray light curve 
is also called `brink' (King \& Ritter 1998).
The presence of a luminosity brink was reported in several Aql X-1 outbursts (Campana et al. 2013).
Campana et al. (2013) found a variable outer disc radius (factor of $\sim 5$), resulting in different
brink luminosities on an outburst by outburst basis.
In addition to the X--ray luminosity brink, Asai et al. (2013) claimed a further  transition in the decay light curve 
of Aql X-1 and 4U 1608--52, which they attribute to the propeller effect (no predictions for the outburst decay 
in case of a propeller effect have ever been made). 
In the case of Aql X-1 they estimated a luminosity of $\sim 8\times 10^{35}\ergs$ at propeller onset.

The first part of the 2010 outburst of Aql X-1, from the peak onward, can be successfully modeled with a Gaussian.
A knee is present at a luminosity of $5.9^{+0.2}_{-0.7}\times 10^{36}$ erg s$^{-1}$, followed by an  exponential decay  with an 
$e-$folding time of $1.7\pm0.1$ d.
While the light curve before the knee could also be described with an exponential, a linear fit after the knee, as predicted in the
disc instability model if we interpret the knee as the brink luminosity,  is clearly rejected by the data.
In addition, the data do not reveal any departure from an exponential decay, at variance with Asai et al. (2013).
On these basis we consider unlikely that the steep decay is due to an outer disc transition. 
This also indicates that outburst light curves probed by all-sky monitoring instruments are difficult to model when the transition 
is close to the instrumental detection limit. 

The 1997 outburst of Aql X-1 was monitored with sufficient detail to estimate the knee luminosity (Campana et al. 1998b). 
By considering the uncertainties involved by the different X--ray instruments, the 1997 and 2010 knee luminosities were 
very similar and statistically compatible, with a mean 0.5--10 keV luminosity of $(5.5\pm0.3)\times 10^{36}$ erg s$^{-1}$.
Given the variations observed, e.g., in the brink luminosity of other outbursts of Aql X-1 (Campana et al. 2013), it is puzzling that the 
same knee luminosity is observed in the two best-studied outbursts (in addition a number of outbursts show a brink luminosity consistent 
with this value). One possibility is that the propeller effect due to the neutron star magnetic field 
plays a role, through the action of the centrifugal barrier. Assuming a relation for the onset of the propeller 
linking the knee luminosity to the neutron star spin period and magnetic field as in Asai et al. (2013),  one can estimate  
a magnetic field of $B\sim (1-4)\times 10^8$ G (including a factor of two uncertainty in the Alfv\`en radius definition).
Thus we are observing the shut-off of the accretion process onto the neutron star surface. 
One possibility is that a ``competition" occurs among the  propeller onset and the (variable) brink onset, depending on the outburst 
evolution (e.g. irradiation properties, mass inflow rate, etc.). In the case of black holes, or (low-magnetic field) white dwarfs, only the 
disc drives the luminosity drop off.  This conjecture can be tested by doing similar in depth studies of the decay behavior in black hole transients.

 \subsection{Temperature evolution}

The temperature evolution observed during the light curve decay was quite fast.
The temperature variations could be fitted with an exponential decay with $\tau=2.5\pm0.3$ d.
This is somewhat slower than the luminosity decay ($1.7\pm0.1$ d), but still extremely fast.
In quasi-persistent transients the decay is much longer in comparison, with $\tau\gsim 100$ d.
Recently the return to quiescence of a few classical neutron star transients has been monitored in detail. 
IGR J17480--2446 in Terzan 5 exhibited an 11-week accretion episode (to be compared with $\sim 9$-weeks of the 2010 Aql X-1 outburst) 
and a slow neutron star cooling ($\sim 800$ d; Degenaar et al. 2013b).
XTE J1709--267 showed a very rapid neutron star temperature decrease following a $\sim 2$ months outburst.
In particular, the temperature decreased from $\sim 160$ eV to $\sim 150$ eV in 8 hr and an exponential decay with $\tau\sim 5.6\pm1.9$ d
(Degenaar et al. 2013a).
IGR J17494--3030 showed a very short 10-d outburst. During an XMM-Newton 
observation the neutron star temperature decreased, with an $e-$folding timescale of $\sim 3$ d 
(Armas-Padilla et al. 2013).
Even though the fast decay of XTE J1709--267 has been interpreted as emission from a cooling layer in the neutron star atmosphere 
(Degenaar et al. 2013a), 
the very fast cooling time of IGR J17494--3030 likely hints to accretion as the powering source of the observed X--ray luminosity.
We note that the X--ray spectrum arising from a low-level accreting X--ray source is very similar to that of a cooling 
neutron star atmosphere (Zampieri et al. 1995; Campana, Mereghetti \& Sidoli 1997).
Given that the X--ray luminosity resulting from accretion onto the neutron star surface is expected to track the mass accretion rate, in this picture the 
luminosity drop off should be caused by a drastic reduction of mass accretion rate.
As discussed above, this may result from a low viscosity disc setting in or from the propeller onset.
In the second case we do not expect a sudden and complete halting of the mass inflow rate, but rather a fast drop of 
the X--ray luminosity. One possibility is therefore that we are observing the closure of the centrifugal barrier.
In this case, the cooling of the neutron star surface is overwhelmed by low-level accretion.

A rapid drop off of the X--ray luminosity has also been observed in the quasi-persistent source XTE J1701--462 (Fridriksson et al. 2010).
This source remained active for  $\sim 600$ d and turned off in about 13 d. Its phase exponential decay, with $\tau\sim 1.2$  
d (Fridriksson et al. 2010), has been interpreted as due to the cessation of the accretion process. 
Being heated during a very long outburst, it is not surprising that the neutron star surface emission set in at a high X--ray luminosity 
$\sim 10^{34}$ erg s$^{-1}$ and showed a slow cooling with exponential decay of $\sim 120$ d. 
Quasi-persistent transient sources as well as XTE J1701--462 experienced long outburst durations, which likely brought their crusts
out of thermal equilibrium with the core. The crust reached the equilibrium with the core on a long timescale (years, see Brown et al. 1998
for details). Due to the shorter outburst  durations and the higher quiescent luminosity, this is not observed in Aql X-1.

\subsection {Quiescent level}

As discussed above, the  quiescent state of Aql X-1 after he 2010 outburst did not show any sign of further decay 
(at variance with quasi-persistent low mass transients).
In addition, the quiescent luminosity is slightly lower than that observed immediately following the 1997 outburst (factor $\sim 1.3$).
The 1997 and 2010 outburst fluences differ instead by a factor of $\sim 1.3$ in the opposite sense.
This can be accounted for by the hard luminosity differences and
might indicate that in quiescence we are observing the true core temperature which is in thermal equilibrium with the atmosphere.
In fact, even if the 1997 and 2010 outbursts were among the strongest the Aql X-1 records in terms of fluence, the core temperature 
is thought to be largely unaffected on short (years) timescales, i.e., between different outbursts (Brown et al. 1998;  Colpi et al. 2001).
Being the heat deposited during outbursts small and the core already hot only a small cooling might be expected.
Unfortunately, the quiescence of Aql X-1 is restless, showing flares (ten-fold increase, Coti Zelati et al. 2013; see also Bernardini et al. 2013), 
and this slow cooling may be difficult to probe. 
In addition, the frequent recurrence of the outbursts hampers the possibility to observe the post-outburst behaviour after individual events for 
long time interval. Consistently, Aql X-1 is one of the few sources lying on the prediction of standard cooling (Heinke et al. 2010; 
Wijnands, Degenaar \& Page 2013). 
Differences in quiescent levels following different outbursts could perhaps indicate different levels of quiescent accretion (if at work) 
or different interaction of a turned on millisecond pulsar with the surrounding medium (Campana et al. 1998a; Papitto et al. 2013), both 
likely involving the power law hard spectral component, or a different composition of the neutron star atmosphere at the end of the outburst 
resulting in different observed luminosity for the same core temperature (Brown et al. 1998; Degenaar et al. 2013b).


\section{Conclusions}

We monitored the decay to quiescence of the Aql X-1 2010 outburst.  Our dataset, based on 5 XMM-Newton, 4 Chandra and 2 Swift observations, 
provides the most detailed view of a neutron star binary transient outburst decay. 
The outburst decay shows a pronounced steepening when the X--ray luminosity drops below $5\times 10^{36}$ erg s$^{-1}$ (at a 4.5 kpc distance).
Afterwards, the luminosity drops exponentially with an $e-$folding time of $\sim 2$ d, reducing its luminosity by a factor of $\sim 1,000$ in $\sim 10$ d.
We were also able to track the evolution of the temperature of the soft spectral component characterising the X--ray spectrum. 
The temperature decrease is quite fast too, with an $e-$folding time of $\sim 3$ d.

Comparing the Aql X-1 2010 outburst decay to the 1997 outburst decay (Campana et al. 1998), we find the same knee luminosity 
and the same fast decay. This likely calls for the same mechanism responsible for the outburst turn off.
We suggest that this occurs due to the propeller inhibition of accretion, induced by the fast spinning neutron star magnetosphere.
The steep luminosity decay would then be due to the closure of the centrifugal barrier, with the temperature decay tracking the switch 
off of the mass inflow onto the neutron star surface.

The quiescent luminosities at the end of the 1997 and 2010 outbursts are somewhat different, 
the quiescent neutron star temperatures are instead well consistent with one another.
This suggests that a hot neutron star core is dominating the thermal balance and X--ray luminosity differences 
can be attributed to a different level of the quiescent power law component.

\section{Acknowledgments}
SC and FB acknowledge useful discussions with P. Pizzochero. RW is supported by an ERC starting grant.
ND is supported by NASA through Hubble Postdoctoral Fellowship grant number HST-HF-51287.01-A from the Space Telescope Science Institute.


\begin{thebibliography}{}

\bibitem[]{}
Armas-Padilla, M., Wijnands, R., Degenaar, N. 2013, MNRAS, 436, L89

\bibitem[]{}
Asai, K., Matsuoka, M., Mihara, T., et al. 2013, ApJ, 773, 117

\bibitem[]{}
Bernardini, F., Cackett, E. M., Brown, E. F., D'Angelo, C., Degenaar, N., Miller, J. M., Reynolds, M., Wijnands, R.
2013, MNRAS, 436, 2465

\bibitem[]{}
Brown, E. F., Bildsten, L., Rutledge, R. E. 1998, ApJ, 504, L95

\bibitem[]{}
Cackett, E. M., Wijnands, R., Miller, J. M., Brown, E. F.,  Degenaar, N. 2008, ApJ, 687, L87

\bibitem[]{}
Cackett, E. M., Brown, E. F., Cumming, A., et al. 2010, ApJ, 722, L137

\bibitem[]{}
Cackett, E. M., Fridriksson, J. K.,  Homan, J.,  Miller, J. M., Wijnands, R. 2011, MNRAS, 414, 3006

\bibitem[]{}
Campana, S., Mereghetti, S., Sidoli, L. 1997, A\&A, 32, 783

\bibitem[]{}
Campana, S., Stella, L., Mereghetti, S., Colpi, M., Tavani, M., Ricci, D., Dal Fiume, D., Belloni, T. 1998a, ApJ, 499, L65

\bibitem[]{}
Campana, S., Colpi, M., Mereghetti, S., Stella, L., Tavani, M. 1998b, A\&A Rev, 8, 279

\bibitem[]{}
Campana, S., Stella, L. 2003, ApJ, 597, 474

\bibitem[]{}
Campana, S., Stella, L., Kennea, J. A. 2008, ApJ,  684, L99

\bibitem[]{}
Campana, S. 2009, ApJ, 699, 1144

\bibitem[]{}
Campana, S., Coti Zelati, F., D'Avanzo, P. 2013, MNRAS, 432, 1695

\bibitem[]{}
Chevalier, C., Ilovaisky, S. A., Leisy, P., Patat, F. 1999,  A\&A, 347, L51


\bibitem[]{}
Colpi, M., Geppert, U., Page, D., Possenti, A. 2001, ApJ, 548, L175

\bibitem[]{}
Coti Zelati, F., Campana, S., D'Avanzo, P., Melandri, A. 2013, MNRAS, in press (arXiv 1312.2379)

\bibitem[]{}
Degenaar, N., Wijnands, R., Wolff, M. T., et al. 2009, MNRAS, 396, L26

\bibitem[]{}
Degenaar, N., Wolff, M. T., Ray, P. S., et al. 2011, MNRAS, 412, 1409

\bibitem[]{}
Degenaar, N., Wijnands, R., Miller, J. M. 2013a, ApJ, 767, L31

\bibitem[]{}
Degenaar, N., Wijnands, R., Brown, E. F., et al. 2013b, ApJ, 775, 48

\bibitem[]{}
Gilfanov, M., Revnivtsev, M., Sunyaev, R., Churazov, E. 1998, A\&A, 338, L83

\bibitem[]{}
Galloway, D. K., Muno, M. P., Hartman, J. M., Psaltis, D., Chakrabarty, D. 2008, ApJS, 179, 360

\bibitem[]{}
Ferrigno, C., Bozzo, E., Falanga, M., et al. 2011, A\&A, 525, A48

\bibitem[]{}
Fridriksson, J. K., Homan, J., Wijnands, R., et al. 2010, ApJ, 714, 270

\bibitem[]{}
Heinke, C. O., Rybicki, G. B., Narayan, R., Grindlay, J. E. 2006, ApJ, 644, 1090

\bibitem[]{}
Heinke, C. O., Altamirano, D., Cohn, H. N., et al. 2010, ApJ, 714, 894

\bibitem[]{}
Homan, J., Fridriksson, J. K., Jonker, P. G., Russell, D. M., Gallo, E., Kuulkers, E., Rea, N., Altamirano, D. 2013, ApJ, 775, 9

\bibitem[]{}
Jonker, P. G., M\'endez, M., Nelemans, G., Wijnands, R., van der Klis, M. 2003, MNRAS, 341, 823

\bibitem[]{}
Jonker, P. G., Campana, S., Steeghs, D., Torres, M. A. P., Galloway, D. K., Markwardt, C. B., Chakrabarty, D., Swank, J. 2005, MNRAS, 361, 511

\bibitem[]{}
Kato, T., Nogami, D., Matsumoto, K., Baba, H. 2004, PASJ, 56, 109

\bibitem[]{}
King, A. R., Ritter H. 1998, MNRAS, 293, L42

\bibitem[]{}
Krimm, H. A., Markwardt, C. B., Deloye, C. J., et al. 2007, ApJ, 668, L147

\bibitem[]{}
Lasota, J.-P. 2001, NewAR, 45, 449

\bibitem[]{}
Masetti, N., Frontera, F., Stella, L., et al. 2000, A\&A, 363, 188

\bibitem[]{}
Matsuoka, M., et al. 2009, PASJ, 61, 999

\bibitem[]{}
Papitto, A., D'A\`i\,ÊA., DiÊSalvo, T., et al. 2013, MNRAS, 429, 3411

\bibitem[]{}
Papitto, A., Ferrigno, C., Bozzo, E., et al. 2013b, Nature, Nature, 501, 517

\bibitem[]{}
Rutledge, R. E., Bildsten, L., Brown, E. F., Pavlov, G. G., Zavlin, V. E. 1999, ApJ, 514, 945

\bibitem[]{}
Rutledge, R. E., Bildsten, L., Brown, E. F., Pavlov, G. G., Zavlin, V. E. 2002, ApJ, 577, 346

\bibitem[]{}
Shahbaz, T., Charles, P. A., King, A. R. 1998, MNRAS, 301, 382

\bibitem[]{}
van Paradijs, J., Verbunt, F., Shafer, R. A., Arnaud, K. A. 1987, A\&A, 182, 47

\bibitem[]{}
Wijnands, R., Guainazzi, M., van der Klis, M.,  M\'endez, M. 2002, ApJ, 573, L45

\bibitem[]{}
Wijnands, R., Homan, J., Miller, J. M.,  Lewin, W. H. G. 2004, ApJ, 606, L61

\bibitem[]{}
Wijnands, R., Rol, E., Cackett, E., Starling, R. L. C., Remillard, R. A. 2009, MNRAS, 393, 126

\bibitem[]{}
Wijnands, R., 2011, proc. of Fast X-ray timing and spectroscopy at extreme count rates (HTRS 2011), Champ\'ery, Switzerland

\bibitem[]{}
Wijnands, R., Degenaar, N., Page, D. 2013, MNRAS, 432, 2366

\bibitem[]{}
Wilms, J., Allen, A., McCray, R. 2000, ApJ, 542, 914

\bibitem[]{}
Zampieri, L., Turolla, R., Zane, S.,  Treves, A. 1995, ApJ, 439, 849

\bibitem[]{}
Zavlin, V. E., Pavlov, G. G., Shibanov, Yu. A. 1996, A\&A, 315, 141

\end{thebibliography}
\end{document}